\documentclass[]{aa}
\usepackage{aas_macros}
\usepackage{mathtools}
\usepackage{siunitx}
\usepackage[normalem]{ulem}

\renewcommand{\vec}{\mathbf}
\usepackage{graphicx}
\usepackage{txfonts}
\usepackage{siunitx}

\usepackage[colorlinks=true, linkcolor=blue, citecolor=blue, filecolor=blue, urlcolor=blue]{hyperref}
\begin{document}
\title{%
Radial and angular evolution of magnetic cloud signatures in the turbulent solar wind: virtual spacecraft analysis%
}
\subtitle{}
\author{%
M. Sangalli\inst{1}%
\thanks{Corresponding author. \email{mattia.sangalli@unifi.it}}%
\and E. K. J. Kilpua\inst{2}
\and A. Verdini\inst{1}
\and S. W. Good\inst{2}
\and J. Pomoell\inst{2}
\and S. Landi\inst{1}%
}
\institute{%
	Dipartimento di Fisica e Astronomia, Università degli Studi di Firenze, Via G. Sansone 1, 50019 Sesto Fiorentino, Italy
	\and
	Department of Physics, University of Helsinki, PO Box 64, 00014 Helsinki, Finland
}
\authorrunning{M. Sangalli {\em et al.}}
\titlerunning{Turbulent magnetic cloud signatures: virtual spacecraft analysis}
\date{Submitted March 17, 2026; accepted June 3, 2026}
\abstract{%
Interplanetary coronal mass ejections (ICMEs) carry magnetic clouds (MCs), large-scale structures with average radial widths about a fifth of an astronomical unit at Earth's orbit. ICMEs display substructures in white light images and reveal rich dynamics across many spatial scales when directly measured by spacecraft. A spacecraft encounter with an ICME can result in smoothly rotating MC intervals or less organised magnetic obstacle (MO) ones.
We investigate how the interplay of expansion, turbulence, and internal cloud dynamics affects magnetic cloud properties, which are reflected in the plasma signatures measured by spacecraft.
We perform high-resolution 2.5D MHD simulations of a magnetic flux rope cross-section, which is embedded in the turbulent, expanding solar wind with the expanding box model. We probe the local plasma properties, and thus the flux rope signatures and angular coherence, with virtual spacecraft.
Our simulations reproduce clear and stable MC signatures when the flux rope core is intercepted by virtual spacecraft. Disordered MO signatures appear at the edges of the flux rope, and are attributed to both expansion and turbulent transport.
We vary some key physical parameters of the flux rope and the environment to understand their effect on the observed coherence and signatures.
The pace of the expanding flow controls the angular extent of MC signatures, whereas the intensity of interplanetary turbulence controls how asymmetric and distorted the flux rope appears at $1$ AU.
The geometry of spacecraft encounters determines whether MC or MO signatures are observed.
The presence of a magnetic structure which can result in MO signatures is strongly controlled by the flux rope's initial (or early) magnetic configuration: MO signatures can only be observed when the axial flux rope field is spatially not well confined by the rope's own magnetic tension, and disappear otherwise.
}
\maketitle
\nolinenumbers
%
%
\section{Introduction}
Coronal mass ejections (CMEs) are disruptive events in the solar atmosphere that cause chromospheric and coronal plasma and magnetic flux to be launched into the heliosphere \citep[e.g.,][]{webb2012LRSP....9....3W}; the magnetic structure of the ejected plasma is usually considered to be a magnetic flux rope (FR), that is, an ensemble of field lines coherently twisted around a common axis, with both ends attached to the Sun \citep[][]{vrsnak1992AnGeo..10..344V, chen1996JGR...10127499C, chen2017PhPl...24i0501C, green2018SSRv..214...46G}. Magnetic flux rope signatures are also routinely observed in interplanetary space, where they have been historically referred to as `magnetic clouds' \citep[MCs,][]{burlaga1981JGR....86.6673B} and later linked to solar eruptions \citep[][]{burlaga1982GeoRL...9.1317B}; typical magnetic cloud signatures include an enhanced magnetic field magnitude, a smooth rotation of the magnetic field vector, and a depressed proton temperature, resulting in a low plasma beta.

Decades of in-situ observations have pointed out the high complexity and rich dynamics of magnetic clouds, which also resulted in nontrivial schematics to capture all their fundamental features \citep[see e.g.][and references therein]{zurbuchen2006cme..book...31Z, wang2018JGRA..123.3238W, al-haddad2025SSRv..221...12A}.
The intrinsic complexity of CMEs is further enhanced by their interplanetary evolution \citep[see e.g. the review by][]{manchester2017SSRv..212.1159M} and by the limitations of a single observation point with respect to a 3D structure; this is reflected in a variety of observed transients, which range from clear magnetic clouds, to more generic `magnetic ejecta' or `magnetic obstacles' \citep[MOs, which present coherent and enhanced magnetic field, but no clear flux rope structure;][]{lepping2008AnGeo..26.1919L, kilpua2009SoPh..254..325K, nieves-chinchilla2018SoPh..293...25N}, to `complex ejecta' \citep[where magnetic signatures are partial or missing, see e.g.][]{burlaga2001JGR...10620957B, nieves-chinchilla2018SoPh..293...25N}.
Magnetic clouds, magnetic obstacles, and complex ejecta might represent different kinds of structures \citep[as suggested, e.g., by][]{gosling1990GMS....58..343G}, may as well arise from interaction of multiple successive CMEs \citep[which seems to be the case for long-duration complex ejecta as argued, e.g., in][]{burlaga2002JGRA..107.1266B}, or may be due to the geometry of the spacecraft crossing \citep[e.g.,][]{jian2006SoPh..239..393J}.
The geometric limitations of single spacecraft measurements have also translated to difficulties in defining minimal sets of signatures that unequivocally define an encounter \citep[see e.g.][]{gosling1990GMS....58..343G, neugebauer1997GMS....99..245N, richardson2004GeoRL..3118804R, zurbuchen2006cme..book...31Z}: no single property is sufficient to declare a CME encounter, and not all the properties are always observed together.

The availability of multi-spacecraft measurements has allowed deeper investigations on the large-scale ICME structure \citep[see e.g.][]{burlaga1981JGR....86.6673B, cane1997JGR...102.7075C, mulligan1999JGR...10428217M, kilpua2011JASTP..73.1228K} and has supported the picture of signatures being dependent on the geometry of the spacecraft encounter.
Recent analyses \citep{davies2020SoPh..295..157D, regnault2024ApJ...962..190R, palmerio2024ApJ...963..108P} have pointed out how even a small change in the spacecraft position might result in different features captured in the observed timeseries.
Multiple spacecraft measurements have also allowed to study the correlation lengths of the ICMEs' magnetic fields, which can be used as a proxy for spatial coherence \citep[see e.g.][]{lugaz2018ApJ...864L...7L, ala-lahti2020JGRA..12528002A, scolini2023ApJ...944...46S, lugaz2024ApJ...962..193L, palmerio2024ApJ...963..108P}.
Moreover, small-scale magnetic field and density structures in coronal mass ejections have been observed and tracked also with heliospheric imagers \citep{hess2023A&A...679A.149H, cappello2024A&A...688A.162C}, showing a range of morphologies which further suggests high spatial complexity.

In addition to the intrinsic properties which make them interesting for fundamental physics, CMEs and MCs are among the major sources of prolonged disturbances in the Earth's magnetic field, which are referred to as geomagnetic storms \citep[see e.g.][]{webb2000JGR...105.7491W, richardson2001GeoRL..28.2569R, huttunen2005AnGeo..23..625H, pulkkinen2007LRSP....4....1P, gopalswamy2022Atmos..13.1781G}.
A better understanding of spatial and temporal variability of these structures is thus also important in the context of space weather, especially on scales below $\sim0.2$ AU in the near-Earth environment \citep[][and references therein]{lugaz2024SSRv..220...73L}.

Numerical simulations and theoretical modelling efforts provide powerful tools to study the magnetic cloud coherence in a controlled environment.
\citet{owens2017NatSR...7.4152O} and \citet{owens2020SoPh..295..148O} have argued that magnetic clouds cease to be coherent structures already from a few tenths of AU, because of the cloud's expansion.
In addition, virtual multi-spacecraft analyses have been applied to numerical simulations of magnetic clouds, both focusing on radial separation \citep{asvestari2021A&A...652A..27A}, and by using a spacecraft swarm \citep{scolini2023ApJ...944...46S}; the latter study \citep{scolini2023ApJ...944...46S} has confirmed the magnetic correlation lengths estimated by in situ observations, and noted how the correlation scales can progressively shrink during propagation, mostly due to the cloud interacting with other large-scale structures.

When captured with high enough cadence, ICME plasma shows clear turbulence signatures \citep[see e.g.][]{leamon1998GeoRL..25.2505L, sorriso-valvo2021ApJ...919L..30S, good2023ApJ...956L..30G}, meaning that a cascade is in place which continuously transfers energy from large to small scales. Moreover, energetic turbulent fluctuations are usually found in the heated and compressed sheath regions in front of fast ICMEs \citep[see e.g.][and references therein]{kilpua2017LRSP...14....5K}.
However, due to computational limitations of most heliospheric numerical models, direct numerical simulations of turbulence in interplanetary magnetic clouds have been carried out only very recently, by \citet{pezzi2024A&A...686A.116P} for a flux rope in a homogeneous plasma and by \citet{sangalli2025A&A...699A.258S} for one in a spherically expanding flow.
The interaction between a magnetic cloud and the turbulent solar wind environment might contribute to erosion or disintegration of the coherent magnetic structure, but might also produce more disordered magnetic signatures, especially at the cloud periphery, where its field is weaker and becomes less organised.
Overall, the impact of turbulence and small-scale structures on magnetic cloud coherence during heliospheric evolution has not been quantified in detail.

In this work, we study high-resolution 2.5D magnetohydrodynamics (MHD) simulations of magnetic clouds (modelled as initially cylindrical flux ropes) which propagate in the heliosphere while interacting with a turbulent background.
We aim to better understand the combined effects of solar wind expansion, interplanetary turbulence, and internal flux rope dynamics on in-situ plasma and magnetic properties: in particular, to what extent turbulence may destroy, obscure, or enhance the typical magnetic cloud signatures.
We analyse the simulations by mimicking multi-point encounters at chosen radial and angular separations, to assess the strength and clarity of several magnetic and plasma signatures, such as enhanced magnetic field magnitude, coherent magnetic field vector rotation, low proton temperature, and low plasma beta.
We further study how variations in physical parameters of both the flux rope, the expanding flow, and turbulence affect their mutual interaction, and, consequently, the spatial extent and quality of coherent signatures.

The structure of this paper is the following. In Sect.~\ref{section-methods}, we briefly describe the numerical model, the assumptions, the setup for virtual in-situ measurements, and the simulations' dataset. In Sect.~\ref{section-results-reference-run}, we present the results of our multi-point analysis of a reference simulation; in Sect.~\ref{section-results-parameters}, we investigate the effect of some key parameters on both the final flux rope shape and on the presence and types of signatures. In Sect.~\ref{section-discussion}, we briefly summarise the results and discuss their value in a broader context.
%
%
\section{Methods}
\label{section-methods}
\subsection{The expanding box model applied to flux ropes}\label{section-methods-ebm}
\subsubsection{Model overview}\label{section-methods-ebm-fr}
We employ 2.5D (2D spatial domain, 3D vector fields) visco-resistive MHD simulations using the expanding box model \citep[EBM,][]{grappin1993PhRvL..70.2190G, grappin1996JGR...101..425G, rappazzo2005ApJ...633..474R}; a description of this approach applied to flux ropes can be found in \citet{sangalli2025A&A...699A.258S}.
In brief, it is assumed that a sufficiently elongated flux rope may be considered locally cylindrical; this allows to limit the simulation to a 2D plane normal to the cylinder axis. We consider a circular flux rope cross-section embedded in a spherically expanding flow to mimic its transport by the background solar wind. In this set-up, spatial gradients and vector field components both decay anisotropically with heliocentric distance, and such trends in turn have a direct impact on the dynamical balance inside the flux rope and at its edges. The flux rope, albeit initially in equilibrium, gets stretched by the spherical flow, expands spherically and radially and eventually relaxes to an expanding equilibrium. All these features are maintained when adding out-of-equilibrium, 2D fluctuations which develop into freely decaying turbulence. Further details and limitations of this modelling choice can be found in \citet{sangalli2025A&A...699A.258S}.
An evident limitation is that there is no large-scale ambient magnetic field. As as result, outside of the flux rope the plasma $\beta$ goes to unrealistically high values, and only approaches unity at the flux rope boundaries.
The simulation code uses periodic boundaries and employs explicit viscosity and resistivity terms, that dissipate small-scale kinetic and magnetic energy and locally heat the plasma as a result; details can be found in Appendix B of \citet{sangalli2025A&A...699A.258S}.

The results of \citet{sangalli2025A&A...699A.258S} showed that turbulence may transport the flux rope's plasma and magnetic field away from its original location; this process occurs preferentially in the non-radial direction because of the loss of magnetic tension due to the (anisotropic) spherical decay of both gradients and magnetic field components with distance.
This effect, together with the natural expansion, increases the possibility to find enhanced magnetic field outside of the flux rope core, which might result in mixed signatures quite far from the flux rope axis.
\subsubsection{Geometry and initial configuration}\label{section-methods-geom-init}
In the EBM, the global Sun-centered spherical coordinates $(r,\Theta,\Phi)$ are rectified into locally cartesian ones ($x$, $y$, $z$), which are transported and stretched by the expanding flow and which correspond at each time to a privileged RTN coordinate system. The flux rope is assumed to be locally cylindrical along its axis, which defines the $z$ direction; $x$ is the local radial coordinate, which always points in the Sun-to-spacecraft direction; finally, $y$ completes the right-handed triad.

The flux rope is modelled as a 2.5D MHD equilibrium.
The magnetic field has cylindrical symmetry and has local cylindrical components $B_z(x,y)$ (axial field) and $B_\theta(x,y)$ (azimuthal field).
The axis is positioned at $(x,y) = (0,0)$ in the simulation plane, and the 2D profiles of both $B_z$ and $B_\theta$ are functions of the local radial coordinate $r' = (x^2 + y^2)^{1/2}$ (distance from flux rope axis) alone.
The axial component $B_z$ peaks at $B_{z,\mathrm{FR}}$ on the axis, and decreases smoothly to zero away from the axis; the width of the $B_z$ profile in the plane is controlled directly by the parameter $\Delta_z$.
The azimuthal component $B_\theta$ is zero on the axis and peaks at $B_{\theta,\mathrm{FR}}$ at a distance from the axis which is controlled by $\Delta_\theta$.
The axial field provides an overpressure, whereas the azimuthal field provides a restoring tension.
The flux rope core is colder than the surrounding environment, which provides stability.
The space inside and outside of the flux rope is populated with 2D pseudo-Alfv{\'e}nic random fluctuations in the $x$ and $y$ components of local velocity and magnetic field.
The flux rope configuration and the procedure for adding fluctuations are described in \citet{sangalli2025A&A...699A.258S}, in Appendix A and Section 2, respectively.
\subsubsection{Parameters and choice of physical units}\label{section-methods-units}
The ideal MHD equations are scale-free, meaning that the physics is independent of the field values, provided that the significant non-dimensional parameters ($\beta$, Mach numbers) are the same.
Here we define an effective plasma beta
\begin{equation}\label{eqn-beta0}
    \beta_0 = 2 \rho_\mathrm{bg} T_\mathrm{bg} / B_\mathrm{FR}^2
    ,
\end{equation}
which represents the ratio between external kinetic pressure and flux rope magnetic pressure. Here $B_\mathrm{FR} = B_{z,\mathrm{FR}} = B_0$ is the initial flux rope peak field (taken as reference magnetic field), $\rho_\mathrm{bg}$ is the proton mass density of both the background and the flux rope (taken as reference density), and $T_\mathrm{bg}$ is the background proton temperature, all in dimensionless units.
The dynamics is controlled by two principal dynamical timescales.

The characteristic Alfv{\'e}n time is defined as $t_\mathrm{A} = L_\mathrm{FR} / c_{\mathrm{A},0}$, where $L_\mathrm{FR} = L_0$ is the initial flux rope diameter (taken as reference length) and $c_{\mathrm{A},0}$ is the Alfv{\'e}n speed obtained from $B_\mathrm{FR}$ and $\rho_\mathrm{bg}$ (taken as reference speed). This is the timescale of Alfv{\'e}n speed propagation and, more generally, of the coherent magnetic dynamics in the flux rope.

The nonlinear time is defined as $t_\mathrm{NL} \sim (k \delta{u})^{-1}$, where $k$ and $\delta{u}$ are typical wave-number and intensity of the turbulent fluctuations. This is the timescale on which turbulent fluctuations act on the plasma.

In the EBM, in addition to the parameters of MHD, we have an additional timescale, the propagation and expansion time, $t_\mathrm{exp} = R_0 / U_0$, where $U_0$ is the constant propagation speed and $R_0$ is the initial heliocentric position.
Thus, an additional nondimensional parameter is introduced: the nondimensional expansion rate
\begin{equation}\label{eqn-exprate}
    \varepsilon_0 = t_\mathrm{A} / t_\mathrm{exp} = ( L_0 U_0 ) / ( c_{\mathrm{A},0} R_0 )
    .
\end{equation}
This parameter controls how fast the expanding flow is, compared to the flux rope internal crossing (Alfv{\'e}n) time. In other words, a low value ($\varepsilon_0 < 1, \ll 1$) means that the internal dynamics of the flux rope is fast compared to the timescales of expansion, that is, the flux rope is `reactive'.

In brief, we adopt the same nondimensionalisation choices for the MHD EBM equations as in \citet{sangalli2025A&A...699A.258S}. Moreover, the starting heliocentric position of the flux rope is fixed as $30 R_\mathrm{sun}$ and the effective plasma beta is always fixed as $\beta_0 = 1$.
As discussed in \citet{sangalli2025A&A...699A.258S}, a lower value of $\beta_0$ would result in an isotropically wider flux rope cross-section at later times in the evolution.

The initial parameters at $R_0 = 30 R_\mathrm{sun}$ can be constrained using statistical arguments:
the flux rope diameter and proton mass density can be extrapolated to $r = R_0$ using the statistical trends derived in \citet{bothmer1998AnGeo..16....1B};
given the spread around the trends, the ranges can be summarised as $L_0 \sim 11 \pm 3 \,R_\mathrm{sun}$ and $N_\mathrm{p} \sim 730 \pm 500 \,\mathrm{cm}^{-3}$ for the flux rope diameter and proton number density, respectively.
The propagation velocity $U_0$ is quite variable for flux ropes, but the employed model applies better to flux ropes whose ejection velocity is close to that of the ambient solar wind (which can be assumed around $300$ to $500 \,\mathrm{km/s}$).
In addition, $U_0$ could be used to limit the possible values of $B_0$, since the two have been found to be positively correlated \citep{pal2018ApJ...865....4P}, namely $V [\mathrm{km/s}] = 44 \, (B [\mathrm{mG}])^{0.74}$ estimated at $r \simeq 10 R_\mathrm{sun}$;
using this relation, $B(V=400\,\mathrm{km/s}) \simeq 1800 \,\mathrm{nT}$.
However, in order to obtain values of $B_0$ at $R_0 = 30 \,R_\mathrm{sun}$, a scaling law has to be assumed for $B(r)$, which is a source of additional uncertainties, as quite many scaling exponents have been reported in literature.

Here we will focus on two values of the nondimensional expansion rate, namely $\varepsilon_0 \simeq 0.4$, which has already been used as a `reference' in \citet{sangalli2025A&A...699A.258S}, and $\varepsilon_0 \simeq 0.8$. For simplicity, these values can be thought to correspond to the following choices:
$L_0 = 9 \,R_\mathrm{sun}$,
$\rho_0 / m_\mathrm{p} = 700 \,\mathrm{cm}^{-3}$,
$B_0 = 350 \,\mathrm{nT}$, thus implying a velocity unit of $c_{\mathrm{A},0} \simeq 290 \,\mathrm{km/s}$. Then $\varepsilon_0 \simeq 0.4$ corresponds to a propagation speed
$U_0 \simeq 380 \,\mathrm{km/s}$, whereas $\varepsilon_0 \simeq 0.8$ corresponds to $U_0 \simeq 760 \,\mathrm{km/s}$.
This particular set of values is naturally arbitrary and other initial parameters corresponding to more or less the same $\varepsilon_0$ value could be explored, as well as different parameters corresponding to different $\varepsilon_0$ values.
It should be kept in mind that the meaning of $\varepsilon_0$ (independently of the physical values that one can choose to match its value) remains that of the ratio between the reference Alfv{\'e}n time, $t_\mathrm{A}$, and the timescale of propagation and spherical expansion, $t_\mathrm{exp}$.
\subsubsection{Simulations dataset description}\label{section-methods-simulations}
\begin{table}
	\caption{Simulations dataset: parameters for different simulation runs.}
	\label{table-runs}
	\centering
	\begin{tabular}{l c c c c c}
		\hline\hline
        run & $\varepsilon_0$ & $t_\mathrm{1AU} / t_\mathrm{A}$ & $B_{\theta,\mathrm{FR}} / B_{z,\mathrm{FR}}$ & $\Delta_\theta / \Delta_z$ & $\delta{B}/B_\mathrm{FR}$ \\
        \hline
        \texttt{A}   & 0.4 & 16 & 1/2 & 1   & 1/4 \\
        \hline
        \texttt{B}   & 0.4 & 16 & 1/2 & 1   & 1/2 \\
        \texttt{C}   & 0.4 & 16 & 1/4 & 1   & 1/4 \\
        \texttt{D}   & 0.8 & 8  & 1/2 & 1   & 1/4 \\
        \texttt{D1}  & 0.8 & 8  & 1/2 & 5/3 & 1/4 \\
        \texttt{F}   & 0.8 & 8  & 1/4 & 1   & 1/4 \\
        \texttt{F1}  & 0.8 & 8  & 1/4 & 5/3 & 1/4 \\
        \hline
	\end{tabular}
	\tablefoot{
		Run A is the reference run. Nondimensional expansion rate $\varepsilon_0$ describes how fast the spherical flow expands compared to the flux rope's typical Alfv{\'e}n time: lower value means reactive flux rope (A, B, C), higher value means fast flow (D, D1, F, F1). Flux rope azimuthal to axial field ratio ($B_{\theta,\mathrm{FR}} / B_{z,\mathrm{FR}}$) is a proxy for the tension; higher means highly twisted flux rope (A, B, D, D1), lower means loosely twisted (C, F, F1). The ratio of the in-plane widths of azimuthal and axial magnetic field components ($\Delta_\theta$ / $\Delta_z$) describes how well the axial field is initially spatially confined inside the azimuthal field, higher ratios meaning better confinement (D1, F1) and lower ratios meaning poorer confinement (A, B, C, D, F). $\delta{B}/B_\mathrm{FR}$ describes energy in fluctuations compared to the flux rope, with run B having higher energy than the other runs.
	}
\end{table}
The main parameters that define our dataset of simulation runs are shown in Table~\ref{table-runs}. All the simulations share the initial domain area $L_x/L_\mathrm{FR} \times L_y/L_\mathrm{FR} = 8\times4$ and the number of grid points $4096\times2048$, and thus the initial spatial resolution of $L_\mathrm{FR}/512$. The actual resolution along $y$ decreases linearly with the heliocentric position of the box $R(t)$, as the box is being stretched. All runs also share the same effective plasma beta $\beta_0 = 2 \rho_\mathrm{bg} T_\mathrm{bg} / B_\mathrm{FR}^2 = 1.0$.
The fluctuations are initialised with a flat omnidirectional power spectral density for both the magnetic field and velocity in the wave-number interval $k_\mathrm{turb} \in [1;4] \times 2\pi/L_\mathrm{FR}$, and their energy is such that on average $\delta{B}/B_\mathrm{FR} = 0.5$ (run B) and $0.25$ (runs A, C, D, D1, F, F1).

Run A will be our reference. With respect to it: run B has double the energy in fluctuations (controlled via $\delta B / B_{\mathrm{FR}}$); run C has half the magnetic tension (controlled via $B_{\theta,\mathrm{FR}} / B_{z,\mathrm{FR}}$); run D has double the nondimensional expansion rate (controlled by $\varepsilon_0$); run F combines the parameters of runs C and D (half magnetic tension and double expansion rate with respect to A); finally, run D1 is equal to run D, but has a narrower $B_z$ profile (controlled by $\Delta_z$), while the width of $B_\theta$ is kept constant, so that the ratio $\Delta_\theta / \Delta_z$ also changes; the same applies for run F1 with respect to run F.

Because of the way in which the initial equilibrium configuration is constructed \citep[see Appendix A of][]{sangalli2025A&A...699A.258S}, a variation in $B_{\theta,\mathrm{FR}} / B_{z,\mathrm{FR}}$ and/or in $\Delta_\theta / \Delta_z$ changes the ratio of magnetic tension to magnetic pressure; in order for the structure to be in equilibrium nonetheless, the temperature profile is adapted accordingly. For example, a lower $B_{\theta,\mathrm{FR}} / B_{z,\mathrm{FR}}$ (as in run C) requires the flux rope core to be colder (with respect to run A) in order for the net force to be zero.
\subsection{Virtual spacecraft and in-situ signatures}\label{section-methods-sc}
\subsubsection{Virtual spacecraft setup}\label{section-methods-sc-setup}
The semi-Lagrangian framework of the EBM allows us to have 2D spatial profiles of magnetic field and plasma parameters at different heliocentric distances, as the box propagates away from the Sun.
For convenience, we will always identify the heliocentric location of each snapshot, $R = R(t)$, with the heliocentric position of the box midpoint, starting at $R(t=0) = R_0 = 30 R_\mathrm{sun}$.
Here we will focus on one-dimensional cuts along the local radial (Sun-to-spacecraft) coordinate, at different heliocentric distances $R$ and over a range of non-radial impact parameters $p$ (or, equivalently, over a range of angular separations $\Delta\alpha$ with respect to the propagation direction).

In the EBM, the box's propagation direction is purely radial: it corresponds to a trajectory $(r,\Theta_0,\Phi_0)$ in some heliocentric spherical reference frame. The arbitrary choice of $(\Theta_0,\Phi_0)$ also fixes the direction of the local (Cartesian) radial coordinate, $x_\mathrm{EBM} \equiv x$. An angular separation in the global reference frame $(\Delta\Theta,\Delta\Phi)$ translates in the local Cartesian frame to a `rectified angular separation', that is, a separation along $y_\mathrm{EBM} \equiv y$ and/or $z_\mathrm{EBM} \equiv z$. In this work, the system is assumed to be invariant along $z$, so the only possible separation is $\Delta y$, which also corresponds to an impact parameter $p$.
We will consider local radial profiles of selected plasma fields at a number of angles $\Delta\alpha$ with respect to the fixed propagation direction $\alpha_0$ (corresponding to $(\Theta_0,\Phi_0)$); in our 2.5D geometry, each angular separation $\Delta\alpha$ then corresponds to a $\Delta y = \Delta y(t) = \Delta y_0 R(t)/R_0$, where $\Delta y_0$ is the separation at $R = R_0$.
With our choice of units (see Sect.~\ref{section-methods-units}) the angular separation between adjacent trajectories used in the following Sections is around $\ang{3}$.

To ease the comparison with actual spacecraft time-series, in the following we will show the local radial profiles, say $f(x)$, with time on the abscissa: this corresponds to the situation where the structure was crossed by a spacecraft moving towards the Sun (negative $V_r$) at exactly the propagation velocity of the box (that is, the ejection velocity of the flux rope), which is assumed to be constant and equal to $U_0$.
The abscissa of one-dimensional cuts will thus be in temporal units given by $t_\mathrm{SC} = - x_\mathrm{EBM} / U_0$.
We will refer to radial cuts with different angular separations $\Delta\alpha$ as different `trajectories'. 

The fact that the EBM reference frame is just a privileged RTN coordinate system heavily simplifies the analysis of the magnetic field vector components along the one-dimensional cuts, even at different angular separations: $\hat{x}_\mathrm{EBM} \equiv \hat{\mathrm{R}}$ points in the purely radial propagation direction, which is also the Sun-spacecraft line; $\hat{z}_\mathrm{EBM} \equiv \hat{\mathrm{N}}$ is precisely along the flux rope axis (which never changes orientation), and $\hat{y}_\mathrm{EBM} \equiv \hat{\mathrm{T}}$ completes the right-handed triad.
In a real configuration, the axis of a propagating flux rope might be tilted with respect to the RTN system, not only in the TN plane, but possibly also in the RN one. The `original' flux rope field components would then be mixed together in the RTN components measured at the spacecraft (as might happen in real spacecraft data), and some signatures or recognisable features might be more difficult to highlight than shown here.

The actual nature (solar latitude, longitude, or a mix of both) of the angular separation $\Delta\alpha$ between adjacent trajectories is rather arbitrary and again depends on how one chooses the orientation of the EBM $z$ coordinate with respect, e.g., to the ecliptic North.
For example, the flux rope elongated structure (its axis) may be thought to lie in the ecliptic plane, so that our local transversal component, $y$, would describe the local latitude: our angular separation $\Delta\alpha$ would then be purely latitudinal. However, as recent studies have pointed out \citep[][and references therein]{weiss2024ApJ...975..169W, al-haddad2025SSRv..221...12A}, the flux rope might be locally distorted, and its axis may be locally oriented in different directions (and by consequence, its cross-section plane would be locally tilted), due to its previous evolution.
\subsubsection{Flux rope, magnetic cloud, and magnetic obstacle}\label{section-methods-sc-signatures}
In the next Sections we will need to identify both the flux rope structure in the simulation plane, and different kinds of transients coming from the single-point virtual spacecraft observations.
The initial flux rope will need to be especially identified at later stages of its evolution, when it may be substantially distorted, so we give the following definitions.
We consider the `flux rope' (FR) to be identified by the largest closed field line in the plane, which we also refer to as the `last closed field line'.
The `flux rope core' is the region inside the last closed field line, whereas the `flux rope edges' refer to the region close to the last closed field line, that is, where magnetic tension drops substantially.

For what concerns the simulated in-situ observations, we consider a `magnetic cloud' (MC) transient to be defined by the combined presence of an enhanced magnetic field magnitude, a smooth rotation of the magnetic field vector, and, additionally, lower than ambient plasma temperature and plasma $\beta$, and higher than ambient total pressure.
We also define a second type of transient, which we name `magnetic obstacle' (MO), as a period of coherent magnetic field enhancement, but with scarce or absent magnetic field rotation and possible absence of other MC signatures.
%
%
\section{Reference run A: global, radial, and angular evolution of signatures}
\label{section-results-reference-run}
%
%
This Section is structured as follows: the reference run A is presented first, both with a 2D overview and by analysing a virtual spacecraft trajectory at successive heliocentric positions. This trajectory has a fixed angular separation $\Delta\alpha = \Delta\alpha^*$ with respect to the initial flux rope propagation direction, and highlights how an MC signature turns into an MO signature, which progressively disappears from the in-situ timeseries as the flux rope resists the spherical expansion.
The final snapshot in time of run A (corresponding roughly to the flux rope being at $R = 1 \,\mathrm{AU}$) is then analysed at different impact parameters $p \equiv \Delta{y}$, that is, different angular separations $\Delta\alpha$ with respect to the purely radial propagation direction.
\subsection{Global magnetic field evolution}
\label{section-results-reference-run_global}
\begin{figure}[]
    \centering
    \resizebox{\hsize}{!}{\includegraphics{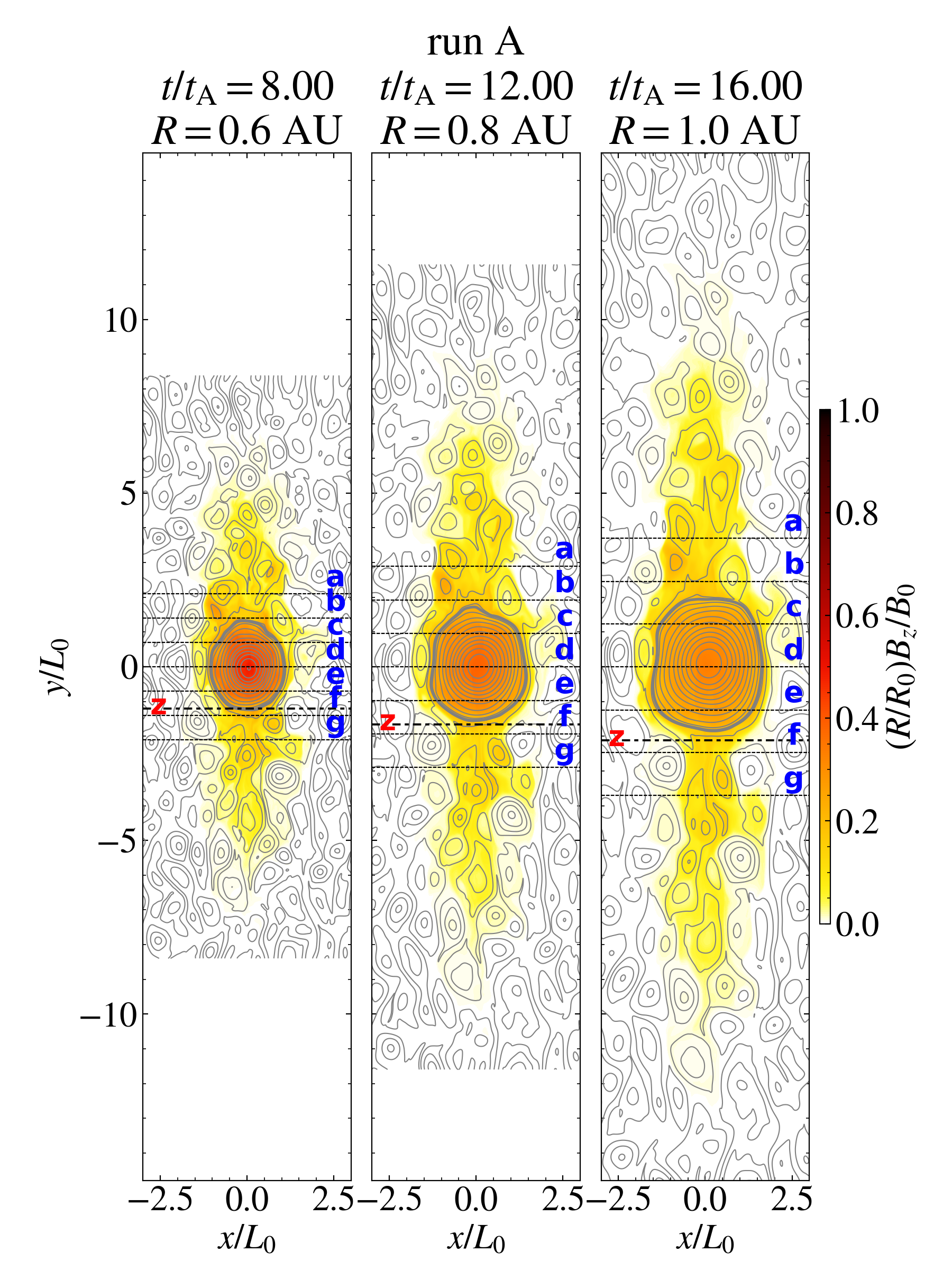}}
    \caption{
        Colour coded intensity map of $B_z$ for run A at increasing heliocentric distances.
        The panels from left to right show the 2D field at successive times, roughly equivalent to $0.6$, $0.8$, and $1.0$ AU. The colour bar has been normalised to $1$ (corresponding to the initial peak value rescaled for the expansion decay). The iso-lines of the out-of-plane magnetic potential $A_z$, which trace the in-plane magnetic field lines, are drawn as grey lines. An estimate of the last closed field line representing the flux rope boundary is drawn as a thicker solid line. The virtual spacecraft trajectories (labelled `a' to `g') are displayed as black dashed lines, and highlight the correspondence between each trajectory and an angle compared to the purely radial direction; an additional trajectory labelled as `z' is displayed as a black dash-dotted line. The distance between the successive snapshots is not in scale. The domain along $x$ is restricted to $x / L_0 \in [-3,+3]$.
    }\label{fig-Bz-runA}
\end{figure}
\begin{figure*}[]
    \centering
    \includegraphics[width=17cm]{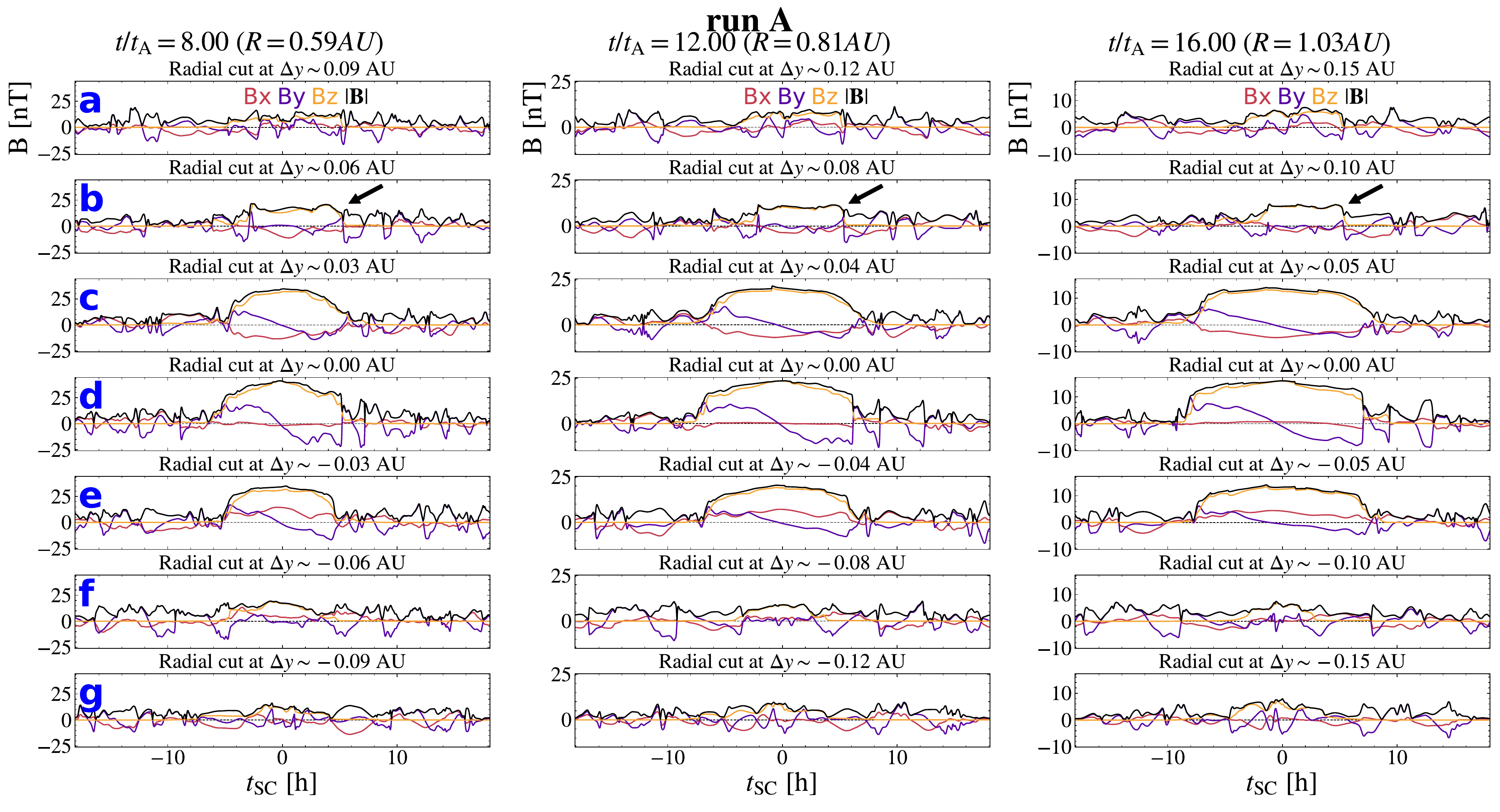}
    \caption{
        Virtual spacecraft time series of the magnetic field components $B_{[xyz]}$ and magnitude $|\vec{B}|$ for run A.
        Each column corresponds to a successive snapshot in time (that is, to a further heliocentric position), with left, centre, and right corresponding to $R\simeq0.6$, $0.8$, and $1.0$ AU, respectively. Each row corresponds to a virtual spacecraft trajectory at a different angle with respect to the radial direction, labelled `a' to `g' (cfr. Fig.~\ref{fig-Bz-runA}). The scale is shared in each column (fixed heliocentric distance), but differs between the columns because of the decay due to expansion. In panel `b' of each column, the feature `O1' is indicated by a black arrow (see text for details).
    }\label{fig-Bxyz-radial-runA}
\end{figure*}
We first focus on the radial and angular evolution of run A.
Figure~\ref{fig-Bz-runA} shows the heliospheric evolution of the axial field $B_z$ (colour coded) with superposed isolines of the out of plane magnetic potential $A_z$, whereas dashed horizontal lines represent the trajectories of the virtual spacecraft encounters (labelled from `a' to `g'). An additional trajectory, labelled `z' and shown with a dash-dotted line, was chosen to highlight some notable features in the radial evolution. The three panels represent increasing heliocentric distance: $0.6$, $0.8$, and $1.0$ AU, from left to right, respectively (not in scale).
Because of the rectification of the spherical geometry in the EBM, a horizontal line corresponds locally to a radial line; moreover, each labelled horizontal line can be traced in time and considered to refer to a given angle $\Delta\alpha$ with respect to some fixed direction (e.g. the initial ejection direction, here considered to be purely radial).
The flux rope core can be identified by both the largest closed field line (an estimate of which is drawn as a thick solid line) and the enhancement in the out-of-plane (axial) field. A substantial increase in the flux rope radial size with distance is also apparent. The axial field dispersion in the transversal in-plane direction ($y$) is highlighted by the lighter shades that extend above and below the flux rope core in the colour coded map; as already discussed and analysed in \citet{{sangalli2025A&A...699A.258S}}, this process is likely due to the combined action of expansion and turbulence.

This 2D evolution already shows that due to the flux rope's expansion, its peak magnetic field value decreases in time (with respect to the assumed spherical decay $\propto 1/R$), while its size increases.
Moreover, the flux rope is slightly deflected from the purely radial path due to chaotic eddies; this implies that at $1$ AU the central trajectory (labelled as `d') is not exactly on-axis, because the structure has been slightly lifted towards the top of the domain (cfr. the central dashed horizontal line in the rightmost panel of Fig.~\ref{fig-Bz-runA}, $R\simeq1$ AU).
The importance of such turbulent deflections might be overestimated here, since we are neglecting the interaction of the flux rope with large-scale bulk flow pertubations such as solar wind streams, and with the large-scale interplanetary magnetic field. Nonetheless, we suggest that the propagation direction of a magnetic cloud which is quietly transported by the solar wind flow might be altered by sufficiently energetic fluctuations (this will be better highlighted in Fig.~\ref{fig-Bz-runsABCDD1}).

The magnetic field components as seen from the virtual spacecraft along trajectories a-g in run A are shown in Fig.~\ref{fig-Bxyz-radial-runA}, where the three columns refer to increasing heliocentric distance from left to right, same as in Fig.~\ref{fig-Bz-runA}.
Clear magnetic cloud signatures (enhanced $|\vec{B}|$, smooth rotation of the field direction) are consistently observed in trajectories `c' through `e', which describe virtual spacecraft trajectories at an angular distance $\Delta\alpha \sim \ang{3}$ from the initial propagation direction, which then correspond to a spatial distance $\Delta y$ of $\simeq0.03$AU, $\simeq0.04$AU, and $\simeq0.05$AU at heliospheric distances of $0.6$, $0.8$, and $1.0$ AU, respectively. The three cuts `c', `d', and `e' are made within the last closed field lines of the flux rope, implying that turbulence does not significantly alter the inner part of the flux rope for run A where fluctuations have relatively low energy ($\delta{B} / B_\mathrm{FR} = 1/4$). 
A coherent radial expansion throughout trajectories c-e can be observed in all the three flux rope magnetic field components, whose duration (i.e. size) increases from the leftmost column ($R\simeq0.6\,\mathrm{AU}$) to the rightmost one ($R\simeq1.0\,\mathrm{AU}$).

On the contrary, trajectory `b' (above the flux rope, $\Delta\alpha \simeq \ang{6}$ from the propagation direction) features a structure with coherent, low variability magnetic field, which however lacks field rotation.
Both spherical expansion and the 2D turbulent transport of the flux rope axial field might account for such signatures. We call this feature from now on `O1'; it is indicated in panels `b' of Fig.~\ref{fig-Bxyz-radial-runA} with a black arrow.

Below the flux rope, along trajectory `f', a brief period of coherent magnetic field with some rotation but weak magnitude is also observed at $0.6$ and $0.8$ AU, though it is not as distinct as in the region above.
The furthest trajectories `a' and `g' do not anymore feature any distinguishable coherent signatures from the background.
%

%
%
Finally, fluctuations are more noticeable and have more energy at small scales at earlier times, which also results in more corrugated flux rope profiles (cfr. for instance trajectory `d' in the middle row of Fig.~\ref{fig-Bxyz-radial-runA}), whereas at later times the fluctuations look smoother, especially inside the flux rope.
This results from freely decaying turbulence, meaning the lack of continuous injection of energy. Consequently, after a few non-linear times, the turbulent energy at the smallest scales gets dissipated and heats the plasma, so that only lower-frequency fluctuations remain \citep[consistently with the analysis presented in][]{sangalli2025A&A...699A.258S}.
\subsection{Evolution of signatures with heliocentric and angular separations}
\label{section-results-reference-run_radial-angular}
\subsubsection{Virtual encounters at a given angle $\Delta\alpha$, at different heliocentric distances}
\label{section-results-reference-run_radial}
\begin{figure*}[]
    \centering
    \includegraphics[width=13cm]{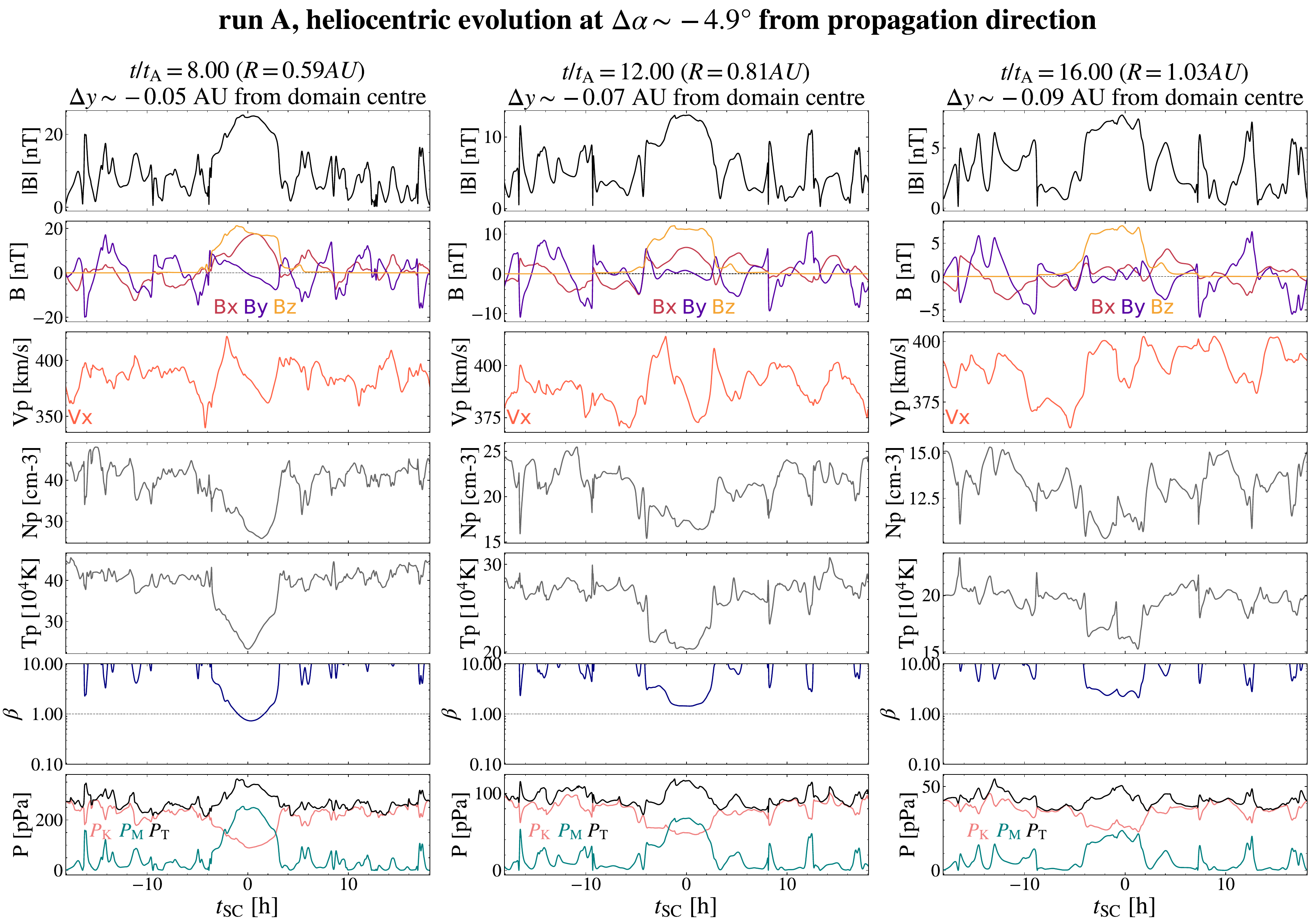}
    \caption{
        Virtual spacecraft time series of magnetic field magnitude $|\vec{B}|$, its components $B_i$, the bulk plasma radial velocity $V_\mathrm{p}$, number density $N_\mathrm{p}$, temperature $T_\mathrm{p}$, beta $\beta$, and pressures (kinetic $P_\mathrm{k}$, magnetic $P_\mathrm{M}$, and total $P_\mathrm{T}$), for run A.
        The columns correspond to a radial cut along trajectory `z' (see Fig.~\ref{fig-Bz-runA}) at successive heliocentric distances, with left, centre, and right corresponding to $R\simeq0.6$, $0.8$ and, $1.0$ AU, respectively. The scales are different between the columns because of the decay due to expansion.
    }\label{fig-fields-insitu-radial-runA}
\end{figure*}
\begin{figure*}[]
    \centering
    \includegraphics[width=13cm]{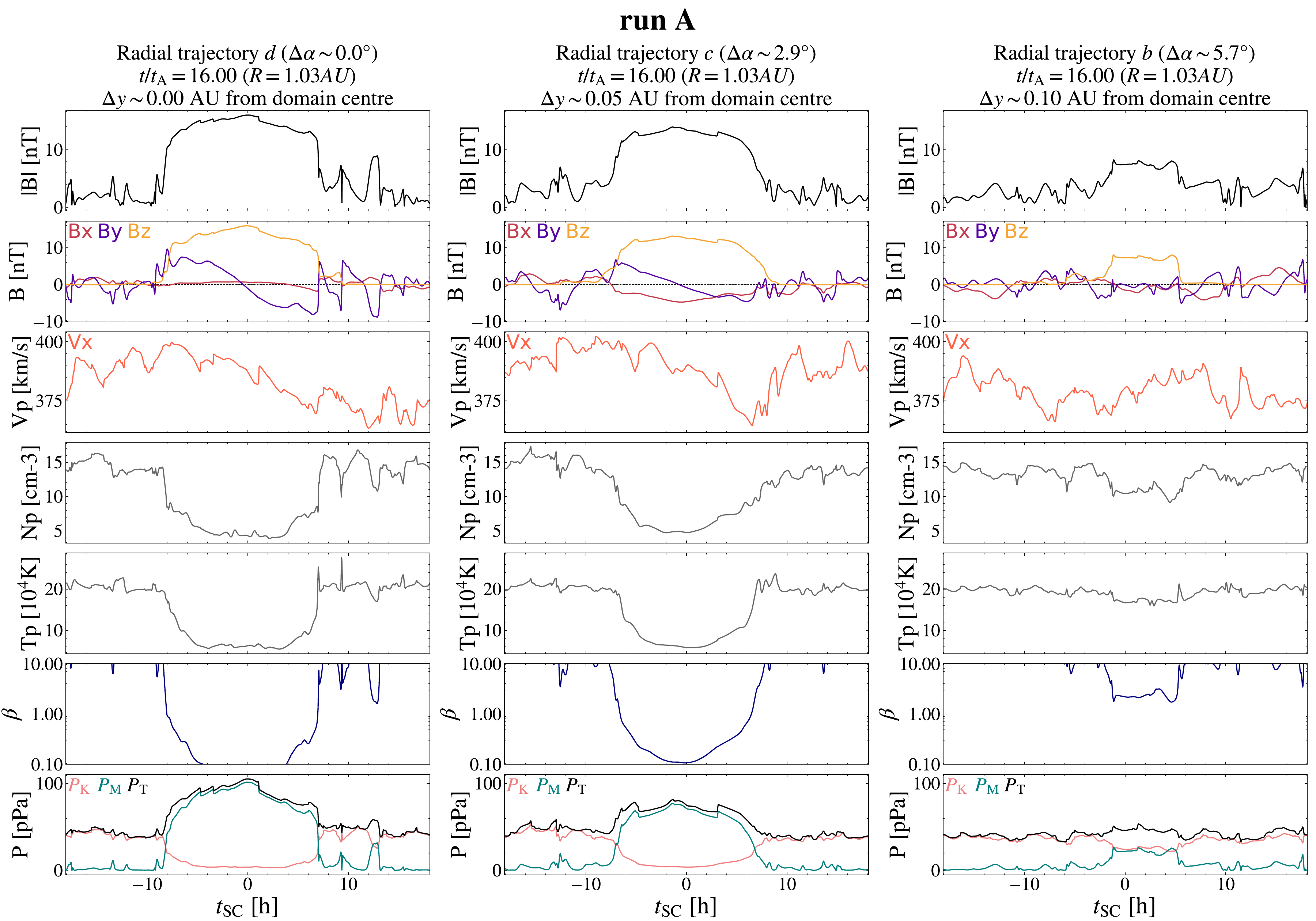}
    \caption{
        Virtual spacecraft time series (same fields as in Fig.~\ref{fig-fields-insitu-radial-runA}), for run A at $R \simeq 1$ AU, along three different virtual spacecraft trajectories.
        The columns correspond to radial cuts along trajectories `d', `c', and `b' (progressively further away from the initial propagation direction), in the leftmost, central, and rightmost columns, respectively. The scales are shared between the columns.
    }\label{fig-fields-insitu-angular-runA}
\end{figure*}
The magnetic field magnitude $|\vec{B}|$, its components $B_i$, the bulk plasma radial velocity $V_\mathrm{p}$, number density $N_\mathrm{p}$, temperature $T_\mathrm{p}$, beta $\beta$, and pressures (kinetic $P_\mathrm{K}$, magnetic $P_\mathrm{M}$, and total $P_\mathrm{T}$) are shown from top to bottom in Fig.~\ref{fig-fields-insitu-radial-runA} along the trajectory labelled as `z', for increasing heliocentric distances from $0.6$ to $1.0$ AU, in the leftmost, central, and rightmost panel, respectively.
Trajectory `z' corresponds to a fixed angular separation $\Delta\alpha = \Delta\alpha^* = \ang{-4.9}$ with respect to the initial propagation direction.
We chose to highlight this particular angular separation because at $R = 0.6$ AU trajectory `z' samples the flux rope edge, just inside the last closed field line, whereas at $R = 1.0$ AU it samples the region just outside of the flux rope core (cf. the dash-dotted trajectory line and the `last closed field line' estimate in the leftmost and rightmost panels of Fig.~\ref{fig-Bz-runA}, respectively).
Coherent magnetic cloud signatures are present for a brief interval in the left panels of Fig.~\ref{fig-fields-insitu-radial-runA} ($R = 0.6 \,\mathrm{AU}$, $\Delta y \sim -0.05 \,\mathrm{AU}$); a smoothly rotating $\vec{B}$, which traces the flux rope lower boundary (cfr. thick dash-dotted reference line in the leftmost panel of Fig.~\ref{fig-Bz-runA}), a coherent head-tail velocity gradient (radial expansion), lower than ambient density and temperature, low $\beta$, anti-correlated magnetic and kinetic pressure profiles, and an enhanced total pressure.

In the central panels of Fig.~\ref{fig-fields-insitu-radial-runA} ($R \sim 0.8 \,\mathrm{AU}$, $\Delta y \sim -0.07 \,\mathrm{AU}$) the smooth $\vec{B}$ rotation becomes less distinct, while the low $\beta$, the radial velocity gradient and a somewhat enhanced total pressure remain visible.
In the rightmost panels of Fig.~\ref{fig-fields-insitu-radial-runA} ($R \sim 1.0 \,\mathrm{AU}$, $\Delta y \sim -0.09 \,\mathrm{AU}$) the magnetic cloud signatures have further weakened. The more ordered $B_z$ interval is, however, still visible and coincident with small decreases in plasma density, temperature and $\beta$.

The region probed by the spacecraft along trajectory `z' displays gradually weaker MC signatures because run A has a rather low nondimensional expansion rate $\varepsilon_0 = 0.4$, which makes the flux rope resist the spherical stretching along the non-radial direction (i.e., the flux rope stays as circular as possible despite the spherical flow trying to stretch it linearly with distance). The FR transversal size expands less than an equivalent angular size $\Delta\alpha^*$ naturally does; thus, the coherently twisted flux rope core progressively `retracts' from the fixed-angle trajectory `z', so that the spacecraft at $R = 0.6$ AU samples a region near the last closed field line, but at $R = 1.0$ AU ends up sampling a region outside the last closed field line.

\begin{figure*}[t]
    \centering
    \includegraphics[width=16cm]{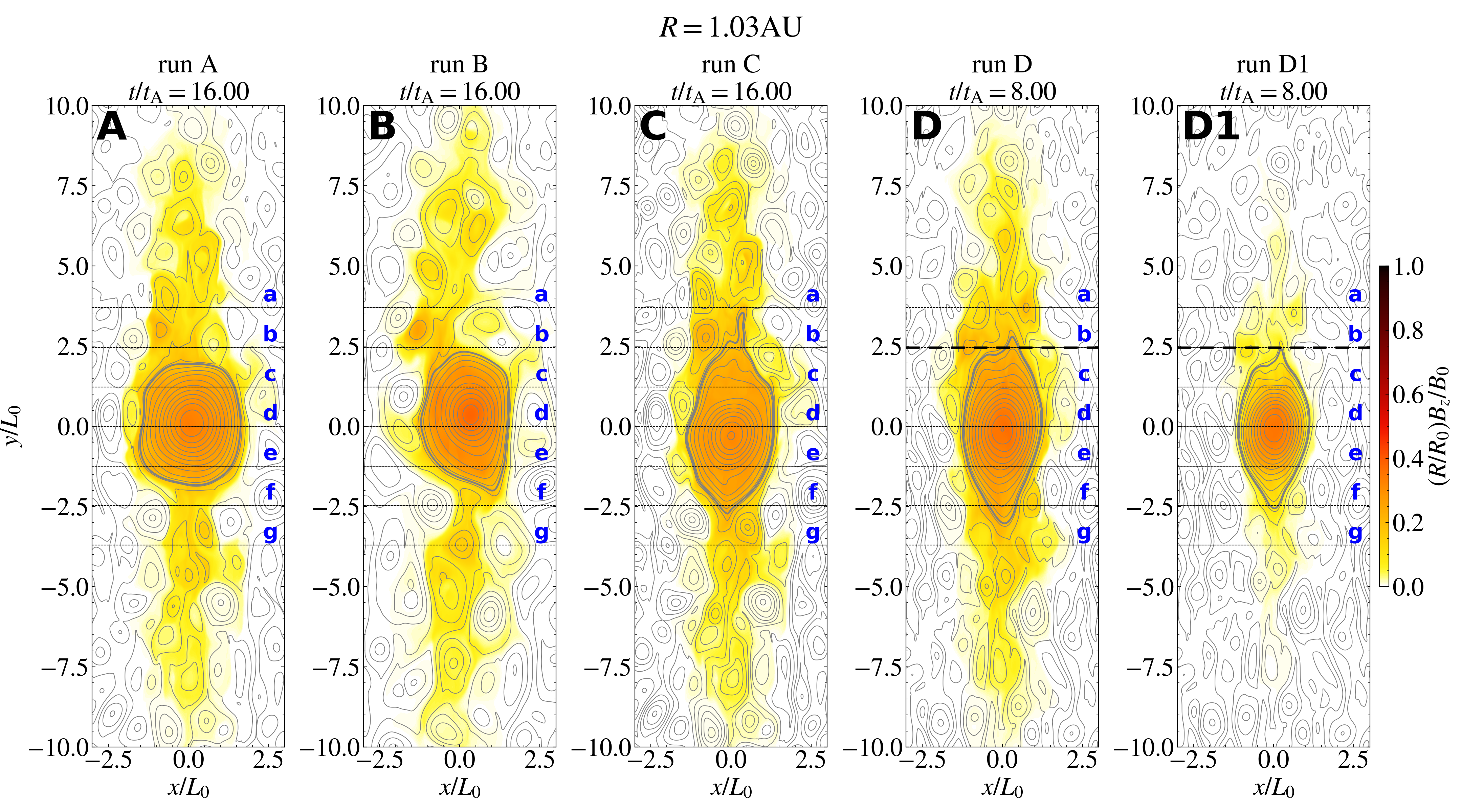}
    \caption{
        Colour coded intensity map of $B_z$, with superposed in-plane magnetic field lines and virtual spacecraft trajectories a-g, with the same conventions as in Fig.~\ref{fig-Bz-runA}, for runs A, B, C, D, and D1, from left to right. Each run is considered at the same heliocentric position, $R \simeq 1$ AU, which corresponds to different times in unit of the reference $t_\mathrm{A}$ (cfr. Table~\ref{table-runs}). In each panel, an estimate of the last closed field line representing the flux rope boundary is drawn as a thick grey solid line. Trajectory `b' is highlighted with a thicker dashed line for runs D and D1. The domain is restricted to $x / L_0 \in [-3,+3]$ and $y / L_0 \in [-10,+10]$.
    }\label{fig-Bz-runsABCDD1}
\end{figure*}

This dynamical flux rope contraction, which leads to the progressive signature loss in the timeseries measured along trajectory `z' at increasing heliocentric positions, would most likely be inhibited by a faster expansion. The magnetic cloud signature caught at $0.6 \,\mathrm{AU}$ would then remain visible in successive encounters with the same $\Delta\alpha^*$.
We will shortly discuss such a scenario when we introduce runs with a higher nondimensional expansion rate in Sect.~\ref{section-results-parameters}, especially in discussing Fig.~\ref{fig-Bz-runsABCDD1}.
\subsubsection{Virtual encounters at a given heliocentric position, at different angles}
\label{section-results-reference-run_angular}
We now focus on how the presence and clarity of the magnetic cloud signatures depend on the angular separation $\Delta\alpha$ at a given distance, namely $R \simeq 1 \,\mathrm{AU}$, corresponding to $t = 16 t_\mathrm{A}$ for run A.
Fig.~\ref{fig-fields-insitu-angular-runA} shows again three columns, with the same fields as in Fig.~\ref{fig-fields-insitu-radial-runA}, but now each column represents a different angular separation with respect to the box centre (i.e., the initial propagation direction): from left to right we show trajectories `d', `c', and `b', which correspond to $\Delta\alpha = \ang{0}, \ang{2.9}, \ang{5.7}$, that is, to a spatial separation $\Delta y = 0$, $0.05$, and $0.1 \,\mathrm{AU}$ from the centre, respectively.
The leftmost panels (traj. `d'), where the flux rope is intercepted nearly on axis ($\Delta\alpha \sim \ang{0}$), exhibit clear magnetic cloud signatures. As the impact parameter increases somewhat in the middle panels (traj. `c', $\Delta\alpha = \ang{2.9}$), the signatures remain clear but become shorter.
In turn, there is a transition to a considerably shorter and fainter MO interval in the rightmost panels (traj. `b', $\Delta\alpha = \ang{5.7}$). This interval features slightly depressed $\beta$ and slightly higher magnetic pressure compared to the surrounding plasma; it corresponds to the feature `O1', which we showed in Fig.~\ref{fig-Bxyz-radial-runA} (rightmost column, panel `b') and discussed earlier in Sect.~\ref{section-results-reference-run_global}. We interpret these weaker signatures as due solely to the axial flux rope field being transported away from the flux rope core by both expansion and turbulence; the cut far from the FR axis allows the spacecraft to measure them.

The decrease in the clarity and intensity of magnetic cloud signatures with increasing angular separation from the flux rope axis is expected, since the flux rope's cross-section has a finite extent. This also implies that the coherent rotation of $\vec{B}$ becomes progressively weaker and less clear as the spacecraft intercepts the flux rope further away from the axis towards its edges. For run A, and with the choice of units discussed in Sec.~\ref{section-methods-units}, the transversal magnetic cloud size may be visually estimated to be around $0.16 \,\mathrm{AU}$ at $R \simeq 1 \,\mathrm{AU}$. The MO signatures shown in the rightmost panels of Fig.~\ref{fig-fields-insitu-angular-runA} are thus intercepted at a fractional impact parameter of $\simeq 125 \%$ from the axis.
%
%
\section{Dynamical control parameters: flux rope shape and the disappearance of magnetic obstacles}
\label{section-results-parameters}
The effect of key dynamical parameters (turbulence strength, twist/tension, and expansion) is next discussed by comparing run A with runs B, C, and D, to investigate separately the roles of turbulence, magnetic tension, and expansion, respectively.
Then, we introduce another key parameter with run D1: the spatial confinement of the axial flux rope field by magnetic tension in the initial configuration.
Finally, two additional runs, F and F1, are further discussed to better highlight the interplay between the parameters that we introduce, especially for what concerns the role of turbulence.
\subsection{Effect of turbulent energy, magnetic tension, and expanding flow}
\label{section-results-parameters_turb-tension-exp}
Figure~\ref{fig-Bz-runsABCDD1} displays the colour coded 2D map of the out-of-plane field $B_z$ together with the isolines of the out of plane magnetic potential $A_z$, which trace the in-plane magnetic field lines. Runs A, B, C, D, and D1 are displayed from left to right, all at $R \simeq 1\,\mathrm{AU}$. As in Fig.~\ref{fig-Bz-runA}, in each panel we highlight an estimate of the last closed field line with a thicker solid line.

%
%
We first consider run B, which differs from run A by  having twice the energy content in fluctuations.
The higher level of fluctuations has significant effects on the overall evolution of the plasma in the simulation domain: turbulent eddies are stronger in field magnitude (not shown) and slightly larger compared to those in run A (cfr. panel A and panel B of Fig.~\ref{fig-Bz-runsABCDD1}).
First of all, the resulting turbulence is more energetic, which means that it distorts and displaces the flux rope more easily and effectively. This is clearly seen in run B from the less symmetric and less circular shape that the flux rope displays at $R \simeq 1$ AU, and from its small deflection towards $y > 0$.
This also implies a lower coherence between time series of plasma parameters observed along the transversal (i.e. angular) direction.
At the same time, the more energetic fluctuations have a shorter typical non-linear time (roughly by a factor $2$) than for run A, and therefore turbulence decays faster compared to the expansion timescale. The resulting fluctuations at later times appear more `washed out' since most of the turbulent energy has already been dissipated.

Run C has the same $\delta{B}/B_\mathrm{FR}$ as run A, but now the peak of azimuthal flux rope field $B_{\theta,\mathrm{FR}}$ (controlling the amount of tension and twist) is reduced by half, while $B_{z,\mathrm{FR}} \equiv B_\mathrm{FR}$ is left unchanged.
This has two notable effects: first, the action of magnetic tension is slower by a factor of two; second, the intensity of the flux rope's coherent boundary sustained by magnetic tension, which resists fluctuations, is halved.
The macroscopic effect of this parameter variation is clearly seen by comparing panels C and A of Fig.~\ref{fig-Bz-runsABCDD1}: run C has a more elliptical and less compact cross-section than run A. This is because the restoring effect of tension acts on slower timescales with respect to expansion, and thus the flux rope gets stretched more easily by the spherical flow. This implies that trajectory `f' now cuts inside the last closed field line (i.e. within the flux rope), and thus is expected to display magnetic cloud signatures; the upper flux rope boundary is also closer to trajectory `b' compared to run A.
In addition, the weaker resistance to chaotic distortions and displacements allows for stronger deformations and deflections. The flux rope cross-section in run C is more irregular than for run A, and more similar to run B.
Similar to run A and B, it has also been deflected, but towards the negative $y$.

Run D has the same parameters as run A, but twice the spherical expansion rate $\varepsilon_0$; this means that the flux rope is embedded in a flow which is faster by a factor of two in units of the internal Alfv{\'e}n time; an obvious consequence is that the structure reaches $1 \,\mathrm{AU}$ in half the Alfv{\'e}n times required for run A ($8$ instead of $16$).
Both flux rope and turbulence experience a stronger stretching, which results in a more elliptical cross-section for the former, and more elongated and less coherent eddies for the latter (this in contrast to run C, where turbulence is not stretched faster, and the eddies are rounder, cfr. panels C and D of Fig.~\ref{fig-Bz-runsABCDD1}).
In run D, the flux rope, thanks to the fairly strong and coherent boundary against turbulence, is only marginally displaced, and experiences only little distortion and erosion by turbulence.
The stronger spherical stretching increases the span of angular separations $\Delta\alpha$ at which radially intersecting spacecraft trajectories can find magnetic cloud signatures, such as trajectory `f' and slightly below trajectory `b', similarly to the case in run C.
%
%
\subsection{Effect of initial configuration: confinement of the axial flux rope field}
\label{section-results-rbz-rbt}
\begin{figure}[]
    \centering
    \resizebox{\hsize}{!}{\includegraphics{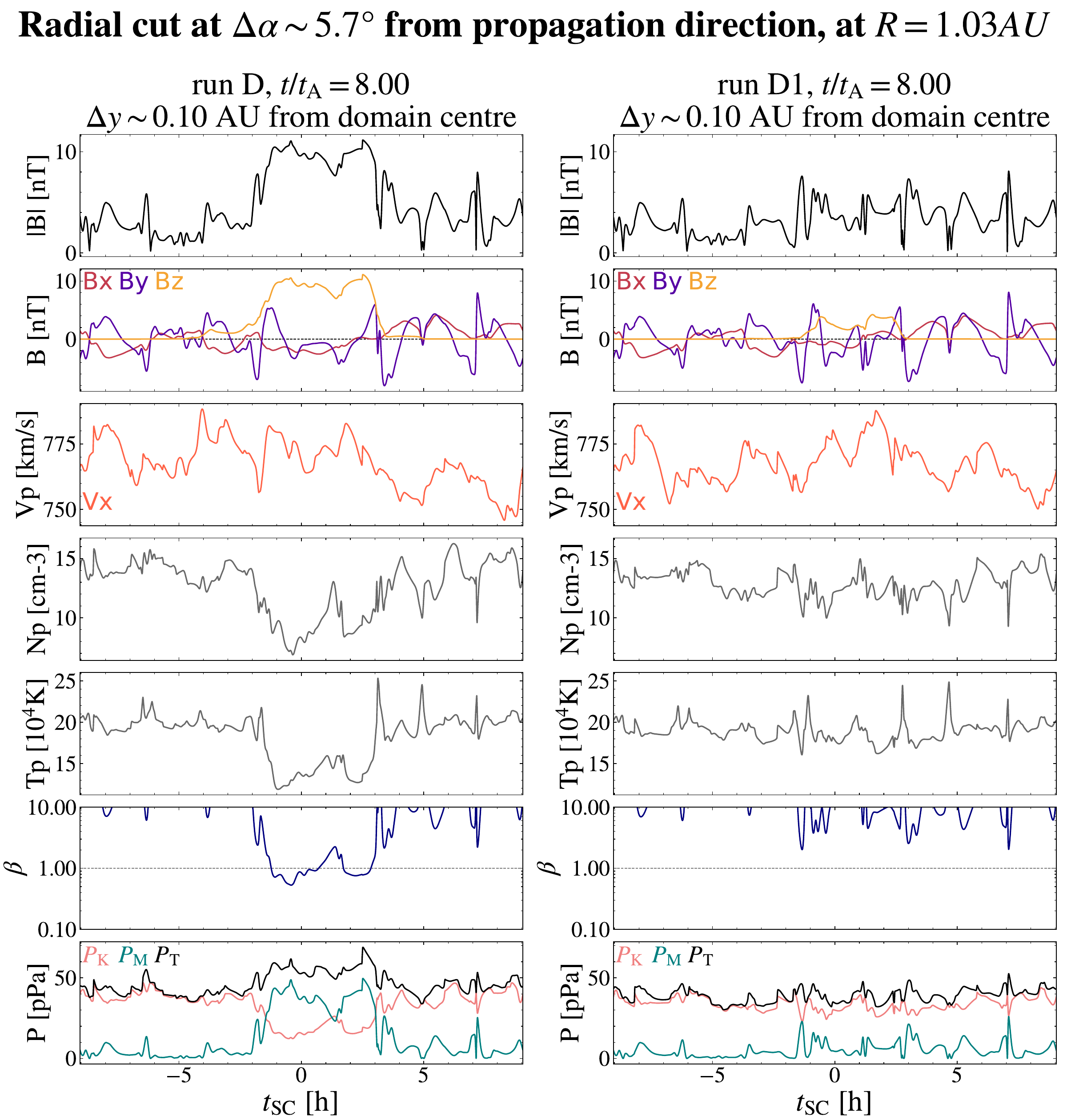}}
    \caption{
        Virtual spacecraft time series (same fields as in Fig.~\ref{fig-fields-insitu-radial-runA}), at $R \simeq 1 \,\mathrm{AU}$, along trajectory `b', for runs D and D1, in left and right columns, respectively.
    }\label{fig-fields-insitu-runD-D1}
\end{figure}
In runs A through D, the initial configuration was such that the axial flux rope field ($B_z$, which is out of the simulation plane) was not totally bounded by magnetic tension (related to $B_\theta$, which lies on the simulation plane).
The main consequence is that the outer portion of $B_z$ does not feel the tension and thus is more free to expand together with the spherical flow, and to be transported around by the turbulent eddies.
To evaluate the importance of this effect on the presence of signatures outside the flux rope core, we consider one additional case:
run D1, with a well bounded axial field as compared to run D.

It is already clear from the 2D overview (cfr. panels D and D1 of Fig.~\ref{fig-Bz-runsABCDD1}) that in run D1 the flux rope stays more compact than in run D; moreover, in run D1 the out-of-plane field $B_z$ (colour coded) is present almost only in the flux rope core region (i.e. inside the last closed field line).
Conversely, in run D (panel D of Fig.~\ref{fig-Bz-runsABCDD1}) the axial field $B_z$ gets much more spread around, and the resulting transversal cross-section extends more than just the size of the flux rope core.

It is interesting to note that the flux rope cross-section has the same shape in both runs, since no dynamical parameters have been changed.
A smaller cross-section for run D1 with respect to run D is expected, as the zone with magnetic pressure ($\sim {B_z}^2$), which makes the flux rope locally expand, is initially narrower.
This effect can be explained as follows. In run D1, the peak axial field $B_{z,\mathrm{FR}}$ is the same as in run D, but confined in a narrower region, which implies that the magnetic pressure gradient associated to the FR axial field is higher in run D1 compared to run D. At the same time, in run D1 the region of influence of magnetic pressure is restricted well inside the region of strong magnetic tension; therefore, the forces that ultimately account for radial expansion are only able to act in a smaller region, and the resulting cross-section is smaller.
The signatures near the flux rope axis (trajectories `c'-`e') are thus only impacted in relation to the lesser radial width of the flux rope.

We now focus specifically on trajectory `b' at $R\simeq1$ AU ($\Delta\alpha = \ang{5.7}$, $\Delta y \simeq 0.1\,\mathrm{AU}$ with respect to the initial propagation direction), for both runs D and D1.
We chose to focus on this trajectory because it samples a region at the flux rope edge for run D, but is well outside of the flux rope core for run D1.
This trajectory is highlighted with a thick black dashed line in the two rightmost panels of Fig.~\ref{fig-Bz-runsABCDD1}.
To better understand the impact of this initial variation on single-point time series at $R\simeq1\,\mathrm{AU}$, Fig.~\ref{fig-fields-insitu-runD-D1} shows the same plasma quantities as in Figs.~\ref{fig-fields-insitu-radial-runA} and~\ref{fig-fields-insitu-angular-runA} along trajectory `b' for both runs D and D1, in the left and right panels, respectively.
It is clear already from the first panel ($|\vec{B}|$) that in run D1 the coherent magnetic enhancement is not present anymore at 1 AU.
The second panel (displaying vector components $B_i$) shows that this comes from the lack of the axial component $B_z$, while the two in-plane fluctuating components $B_x$ and $B_y$ remain almost unchanged compared to run D.
At the same time, the third panel ($V_\mathrm{p}$) shows that run D exhibits a head to tail speed gradient, although quite distorted, whereas run D1 displays only disordered fluctuations.
Both the low density and low temperature intervals, clearly visible for run D, have completely disappeared in run D1. Together with the rather flat magnetic field profile, this results in a $\beta > 1$ interval, almost indistinguishable from the surrounding environment. In contrast, run D has low $\beta$ ($\lesssim 1$) during the same interval.
Finally, also the total pressure, which is slightly higher than the background value for run D, is essentially constant for run D1.
Repeating the same analysis for runs A1, B1, and C1 provides essentially the same result (data not shown).

We may thus conclude that, in the limit of validity of this numerical approach, MO intervals can arise from the in-situ measurements of magnetic structures at or near the FR edges; such structures in turn arise because of the expansion and turbulent transport of the axial FR field which was not initially confined by the FR twisted field.
The magnetic structures at the flux rope edges (and thus the in-situ MO signatures) are lost if the flux rope is initially more compact, that is, if the tension effectively confines the axial field at early stages.
\subsection{Additional remarks on the role of dynamical parameters: runs F and F1}
\label{section-results-parameters_extreme}
\begin{figure}[]
    \centering
    \resizebox{\hsize}{!}{\includegraphics{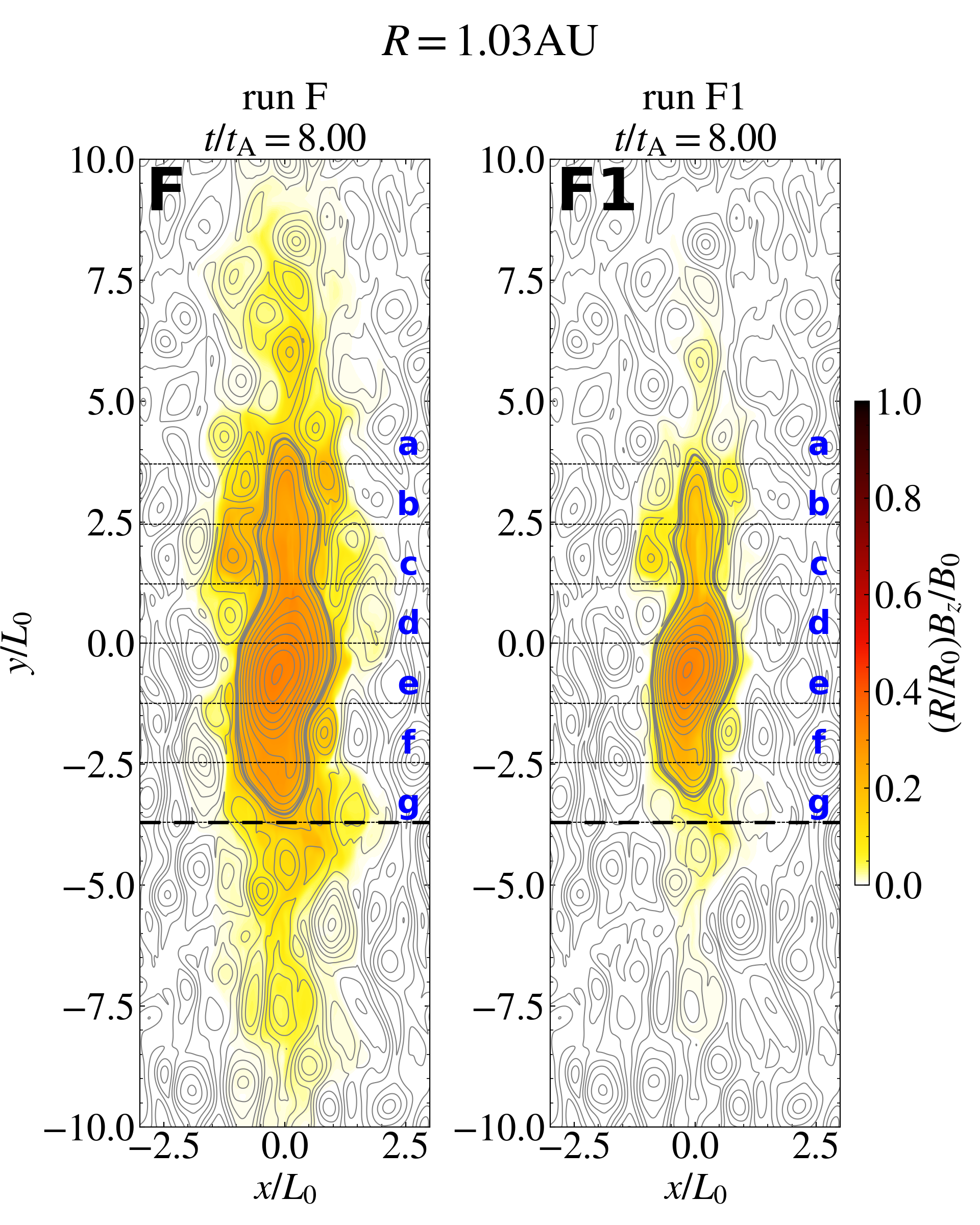}}
    \caption{
        Same as Fig.~\ref{fig-Bz-runsABCDD1} for runs F and F1, in the left and right panels, respectively. In both panels, an estimate of the last closed field line representing the flux rope boundary is drawn as a thick grey solid line. Trajectory `g' is highlighted with a thicker dashed line in both panels.
    }\label{fig-Bz-runsFF1}
\end{figure}
\begin{figure}[]
    \centering
    \resizebox{\hsize}{!}{\includegraphics{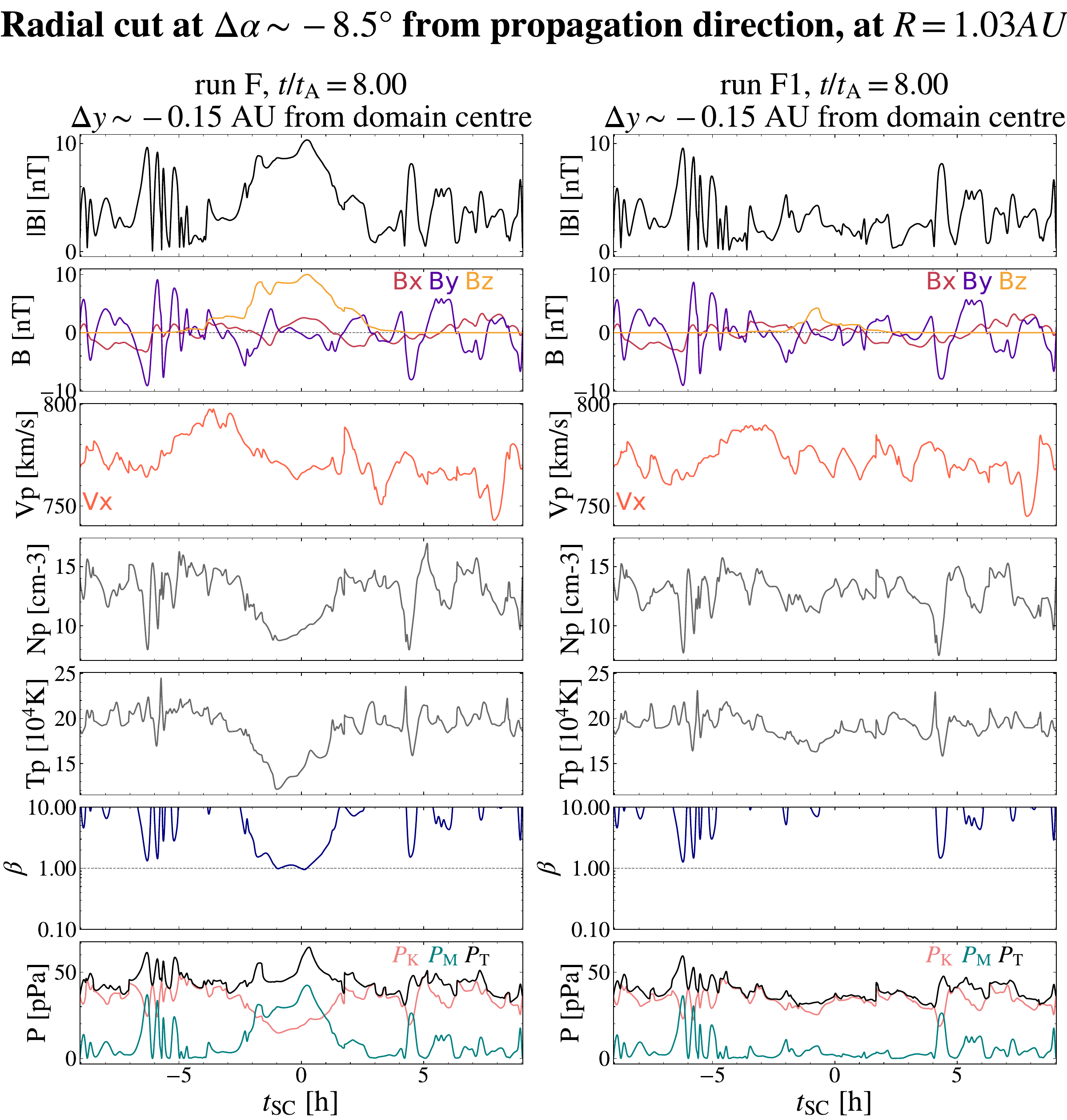}}
    \caption{
        Same as Fig.~\ref{fig-fields-insitu-runD-D1}, here along trajectory `g' for runs F and F1 in the left and right columns, respectively.
    }\label{fig-fields-insitu-runF-F1}
\end{figure}
To further validate our analysis, we also discuss two additional runs.
Run F combines the parameter variations of runs C and D with respect to reference run A. In practice, with respect to run A, run F has a weaker and slower magnetic tension (by a factor of two), and is embedded in an expanding flow which is twice as fast.
With this choice our aim is to investigate a flux rope which is poorly reactive and resistant both to expansion and to turbulent deflections and distortions, in order to understand whether this higher ‘malleability’ can produce MO signatures even when the flux rope axial field is initially well spatially confined by the tension.
We thus also consider run F1, which differs from run F only regarding the initial confinement of the flux rope's axial field component $B_z$ inside the twisted field component $B_\theta$.

The magnetic field configuration at $R \simeq 1\,\mathrm{AU}$ is shown in Fig.~\ref{fig-Bz-runsFF1}, similarly to Fig.~\ref{fig-Bz-runsABCDD1}, with runs F and F1 in the left and right panels, respectively. As in Fig.~\ref{fig-Bz-runsABCDD1}, in both panels we highlight an estimate of the last closed field line with a thicker solid line, and the radial cut of trajectory `g' with a thicker black dashed line.
Similarly to the choice of trajectory `b' in the preceding section, here we chose to focus on trajectory `g' because it samples a region at the flux rope edge for run F, but is well outside of the flux rope core for run F1.
The 2D magnetic field profile of run F is considerably more distorted when compared to all other investigated runs previously discussed and shown in Fig.~\ref{fig-Bz-runsABCDD1}.
The lower coherent resistance related to magnetic tension, together with the strong spherical stretching, makes the flux rope in run F even more susceptible to the action of turbulent eddies than the other explored cases.
A quite consistent deflection towards $y < 0$ is apparent, whereas in the upper domain ($y > 0$), the flux rope's top half has been elongated (along $y$) and compressed (along $x$), resulting in a narrow but angularly extended portion.
A light shade (weak $B_z$) surrounds the flux rope core for run F, but disappears almost totally for run F1, analogously to what happened for run D1 compared to run D, visible in the two rightmost panels of Fig.~\ref{fig-Bz-runsABCDD1}.

Figure~\ref{fig-fields-insitu-runF-F1} represents the cuts for runs F and F1 in the same format as Fig.~\ref{fig-fields-insitu-runD-D1} did for runs D and D1: runs F and F1 are traversed along trajectory `g' at $R\simeq1$AU, corresponding to $\Delta\alpha = \ang{-8.5}$, that is, $\Delta y \simeq -0.15\,\mathrm{AU}$, with respect to the initial propagation direction.
The field profiles highlight a very similar outcome to cases D and D1 (cfr. Fig.~\ref{fig-fields-insitu-runD-D1}), in that the coherent MO signature present for run F disappears for run F1.

On the one hand, this suggests that turbulence is most effective against loosely wound flux ropes which are also effectively stretched by the spherical flow; in those cases, one can observe MC signatures with quite variable duration and field strength depending on where the structure is encountered.
On the other hand, this further example confirms that such enhanced flux rope `malleability' is still not enough to produce MO signatures when the axial flux rope field is well confined by magnetic tension.
%
%
%
\section{Discussion}
\label{section-discussion}
Using virtual spacecraft encounters applied to high-resolution pseudo-Lagrangian 2.5D MHD simulations, we have analysed the heliospheric evolution and angular variability of the magnetic and plasma signatures of an initially coherent flux rope embedded in the turbulent and expanding solar wind. 
We briefly summarise here the main results of this study:
\begin{enumerate}
\item
    spacecraft encounters on or near the flux rope axis result in clear magnetic cloud (MC) signatures, both in magnetic field and in plasma parameters, despite the action of turbulent fluctuations;
\item 
    the flux rope's interaction with the expanding flow and interplanetary turbulence (in particular, the flux rope axial field expansion and its turbulent transport) produce substructures at and near the flux rope edges, that display little or absent coherent rotation and only partial MC properties. When such substructures are encountered by spacecraft at the flux rope edges, they result in magnetic obstacle (MO) signatures;
\item
    a higher ratio $t_\mathrm{A} / t_\mathrm{exp}$ (a faster expansion compared to the flux rope's crossing time) leads to a more elliptical cross-section and, consequently, a greater angular extent of the coherent MC signatures;
\item 
    MC or MO signatures observed in-situ at a given heliospheric distance may be lost when the structure is intercepted at the same angle at a larger distance. This is interpreted to be due to internal non-radial motions related to the flux rope's dynamics \citep[as discussed by][]{sangalli2025A&A...699A.258S};
\item
    the ratios between dynamical timescales typical of magnetic tension, turbulence, and expansion affect the final cross-section shape and the angular variability of in-situ plasma properties; the presence of MC signatures and the existence of MO signatures however are not altered;
\item
    the presence of the magnetic substructures at the flux rope edges, which are responsible for MO signatures, is strongly dependent on the initial confinement of the flux rope axial field by its azimuthal one.
\end{enumerate}
In order to better connect these mostly qualitative results to observations and previous studies, below we report some numerical estimates.
Upon visual inspection, we find that the coherent MC signatures extend in the non-radial direction in the simulation plane (normal to the flux rope axis) for an angle of $\ang{10}-\ang{15}$ (spatial size of $0.15-0.23$ AU at $1$ AU); the range of variation comes from the different simulation parameters.
These values are consistent with the average MC angular extent of $\simeq\ang{12}$, which comes from an average radial width at $1$ AU of $0.21$ AU \citep{lepping2005AnGeo..23.2687L} assuming a circular cross-section.
Our estimates for the non-radial spatial extent (normal to the flux rope axis) naturally increase to about $\ang{13}-\ang{18}$ (that is, $0.2-0.3$ AU) if we also consider the substructures at the flux rope edges that are responsible for the MO signatures.

Our estimated non-radial spatial extent of the magnetic structure at $1$ AU is thus close to the non-radial scale lengths that have been estimated for the total magnetic field in magnetic ejecta near $1$ AU, namely $0.25-0.35$ AU \citep[][]{lugaz2018ApJ...864L...7L}.
However, in our 2.5D simulations we can only probe the magnetic structure's extent in the plane normal to the flux rope axis, and we cannot probe its size along the axis, which most likely leads to a smaller size.

As already stressed in Sect.~\ref{section-methods-units}, the values of our numerical estimates depend on the choice of physical units and on the value of the control parameters.
In particular, we find that the extent of the magnetic cloud's angular coherence depends mainly on the ratio $t_\mathrm{A}/t_\mathrm{exp}$, which controls the final cross-section elongation.

The presence of MO signatures instead of MC ones in the virtual in-situ measurements seems to arise mostly from the geometry of the spacecraft encounter: the closer the spacecraft is to the flux rope core, the stronger and more organised the signature is (with a progression from MC to MO to no signature moving further away from the core).
In this sense, our results are consistent with the interpretation and sketch by \citet{jian2006SoPh..239..393J}, with the notable difference that in this work the `magnetic obstacle' region is given by turbulence and by the transported axial flux rope field which was not initially confined by magnetic tension.

We find that the key factor determining the presence or absence of MO signatures is the initial magnetic flux rope configuration, or, more precisely, how well the flux rope axial field is initially spatially confined by its azimuthal (twisted) field.
Our initial flux rope magnetic properties (that is, at $R = 30 R_\mathrm{sun}$) might be interpreted as the result of an early erosion process, which would cause the azimuthal field to be weakened and the axial field to be more poorly confined. Alternatively, they could be due to the initial coronal flux rope configuration, perhaps based on the release mechanism.

The crucial role of the initial axial flux rope field confinement in producing MO signatures appears consistent with previous observational results.
At solar minimum ICMEs are found to be smaller and weaker than at solar maximum \citep{jian2008SoPh..249...85J, jian2011SoPh..274..321J}; in our framework, where we model ICMEs as flux ropes, this may be interpreted as flux ropes having their axial field which is confined better by their azimuthal one at solar minimum than at solar maximum.
Such a modelling assumption may be justified also by considering the expected increase in magnetic complexity of CMEs towards solar maximum \citep[as suggested e.g. by][]{richardson2004GeoRL..3118804R}.
Using this assumption and considering the results presented in the present work, at solar minimum interplanetary flux ropes would produce either an MC signature or no signature at all, whereas at solar maximum they would feature all three possibilities: an MC, an MO, or no signature.
Such results would thus be consistent with a higher fraction of magnetic clouds among the encountered ICMEs at solar minimum compared to solar maximum \citep[as in e.g.][]{richardson2004GeoRL..3118804R}.

By examining the runs presented in this work (for example, the panels of Figs.~\ref{fig-fields-insitu-radial-runA}, \ref{fig-fields-insitu-angular-runA}, \ref{fig-fields-insitu-runD-D1}, and~\ref{fig-fields-insitu-runF-F1}), we can notice an approximate correspondence between intervals of $\beta < 1$, excess of total pressure, and a recognisable and dominant head-tail radial speed gradient.
Such correspondence between magnetic and velocity signatures appears consistent with previous results for magnetic clouds \citep[e.g.,][]{demoulin2009A&A...498..551D}, even though low-beta clouds that are not expanding are also observed.
This may not always be the case for magnetic obstacle intervals: such correspondence is rather clear in the left column of Fig.~\ref{fig-fields-insitu-runD-D1}, disturbed and less clear in the left column of Fig.~\ref{fig-fields-insitu-runF-F1}, and almost absent in the rightmost column of Fig.~\ref{fig-fields-insitu-angular-runA}.
In the framework used in this work, MO intervals are captured by virtual spacecraft that encounter the flux rope edges: there, the FR axial field (which dominates the pressure and thus the dynamics) is weaker, the beta is more likely to be higher than $1$, and as a result the radial expansion signatures may be obscured and/or neutralised by turbulent eddies.

We also suggest that our results may have a relevance for quasi-real-time space weather forecasting using future space missions orbiting close to Earth on the dayside (sub-L1), such as HENON \citep{cicalo2025Ap&SS.370...83C} and MIIST \citep{lugaz2024SSRv..220...73L}.
If an MO signature is observed by a spacecraft off the Sun-Earth line, an approximate forecast may be issued regarding the possibility of an organised magnetic cloud structure within, say, $\ang{5}-\ang{10}$, thus potentially impacting on Earth.

The results of this study could be further confirmed and improved by using actual in-situ data, possibly coming from multiple spacecraft; this could also help constrain some of the model parameters.
An extension to three-dimensional simulations may also help reproduce richer magnetic obstacle signatures, because of the greater dimensionality.
\begin{acknowledgements}
The authors wish to thank the anonymous referee for their constructive comments.
M.S. acknowledges partial financial support from the European Union - Next Generation EU - National Recovery and Resilience Plan (NRRP) - M4C2 Investment 1.4 - Research Programme CN00000013 ``National Centre for HPC, Big Data and Quantum Computing'' - CUP B83C22002830001. E.K.J.K. acknowledges support from the Research Council of Finland (RCF) Centre of Excellence in Research of Sustainable Space (grant 352850; FORESAIL). S.W.G. is supported by an RCF Academy Fellowship (grants 338486, 346612 and 359914). J.P. acknowledges RCF projects 343581, 364852 and 370793.
A.V. and S.L. acknowledge partial financial support from the European Union – Next Generation EU – National Recovery and Resilience Plan (NRRP) – M4C2 Investment 1.1- PRIN 2022 (D.D. 104 del 2/2/2022) – Project ``Modeling Interplanetary Coronal Mass Ejections'', MUR code 31. 2022M5TKR2, CUP B53D23004860006.
\end{acknowledgements}
\bibliographystyle{aa}
\bibliography{paper}
\end{document}